\input{aipcheck}

\documentclass[amsmath,showpacs,amssymb
    ,final            % use final for the camera ready runs
  ]
  {aipproc}

\layoutstyle{8x11single}

\newcommand\beq{\begin{equation}}
\newcommand\eeq{\end{equation}}
\newcommand\beqa{\begin{eqnarray}}
\newcommand\eeqa{\end{eqnarray}}

\newcommand{\gd}{\dot{\gamma}}

\newcommand{\wa}{\mu}

\newcommand{\wz}{\epsilon}

\newcommand{\text}{\mathrm}

\begin{document}

\title{Does the Chapman--Enskog expansion for viscous granular flows
converge?}


\author{Andr\'es Santos}{
  address={Departamento de F\'{\i}sica, Universidad de
Extremadura, E-06071 Badajoz, Spain} }

\begin{abstract}
This paper deals with the convergence/divergence issue of the
Chapman--Enskog series expansion of the shear and normal stresses
for a granular gas of inelastic hard spheres. {}From the exact
solution of a simple kinetic model in the uniform shear and
longitudinal flows, it is shown that (except in the elastic limit)
both  series  converge and their respective radii of convergence
increase with inelasticity. This paradoxical result can be
understood in terms of the time evolution of the Knudsen number and
the existence of a nonequilibrium steady state.
\end{abstract}

\maketitle

%%%%%%%%%%%%%%%%%%%%%%%%%%%%%%%%%%%%%%%%%%%%
%% MAINMATTER
%%%%%%%%%%%%%%%%%%%%%%%%%%%%%%%%%%%%%%%%%%%%

%\section{<A section>}

%\section{<Another section>}

%\subsection{<A subsection>}

%\subsubsection{<A subsubsection>}

%\paragraph{<A subsubsubsection>}

\paragraph{Introduction}
In a Newtonian fluid the irreversible parts of the stress tensor
$P_{ij}$ are linear in the velocity gradients $\nabla_i u_j$.  For
instance, if the flow is incompressible (i.e.,
$\nabla\cdot\mathbf{u}=0$), $P_{xy}=-\eta_0 {\partial u_x}/{\partial
y}$, where $\eta_0$ is the shear viscosity and it has been assumed
that $\partial u_y/\partial x=0$. Analogously, in a compressible
flow characterized by a rate-of-strain tensor $\nabla_i
u_j=(\partial u_x/\partial x)\delta_{ix}\delta_{jx}$ one has
$P_{xx}=p-(4/3)\eta_0 {\partial u_x}/{\partial x}$, where $p$ is the
hydrostatic pressure and the bulk viscosity has been neglected (as
happens in the low-density limit). In the  case of a dilute gas,
Newton's law  is derived from the Boltzmann equation under the
assumption that the  characteristic hydrodynamic length ($L$)
associated with the gradients of density ($n$), temperature ($T$),
or flow velocity ($\mathbf{u}$) is much larger than the mean free
path $\ell$ of the gas particles, i.e., $\ell \ll L$. More in
general, the Chapman--Enskog (CE) method expresses the solution of
the Boltzmann equation as an expansion in powers of the Knudsen
number $|\wa|\equiv \ell/ L$ \cite{CC70}. The leading terms in the
CE expansion yield the Navier--Stokes constitutive equations and
provide explicit expressions for the transport coefficients (like
the shear viscosity $\eta_0$).

An important basic question is the nature (convergent versus
divergent) of the CE expansion.  To isolate the problem, let us
consider the following  subclasses of the full CE series:
\beq
P_{xy}=-\sum_{k=0}^\infty \eta_k \left({\partial u_x}/{\partial
y}\right)^{2k+1},\quad P_{xx}=p-\frac{4}{3}\sum_{k=0}^\infty \eta_k'
(\partial u_x/\partial x)^{k+1},
\label{2}
\eeq
where $\eta_0=\eta_0'$ is the shear viscosity, $\eta_1'$ is a
Burnett coefficient, $\eta_1$ and $\eta_2'$ are super-Burnett
coefficients, and so on. The two full CE series of $P_{xa}$ (with
$a=y$ and $a=x$, respectively) reduce to the partial series
\eqref{2} if (and only if) $\nabla_i u_j=\gd_{xa}
\delta_{ia}\delta_{jx}$ and $\nabla n=\nabla
T=\nabla\gd_{xa}=\mathbf{0}$, i.e., the only nonzero hydrodynamic
gradient is a uniform shear or longitudinal rate $\gd_{xa}\equiv
\partial u_x/\partial a$. These are precisely the conditions characterizing two
well-defined physical states: the  uniform shear flow (USF, $a=y$)
\cite{LE72,GS03,SBD86} and the uniform longitudinal flow (ULF,
$a=x$) \cite{GK96,S00}. In both states the  characteristic
hydrodynamic length
 is  $L\sim
\sqrt{2T/m}/|\gd_{xa}|$ (where $m$ is the mass of a particle and the
temperature is measured in energy units), so that the Knudsen number
becomes the absolute value of $\wa\equiv\gd_{xa}/\nu$, where
$\nu\sim \sqrt{2T/m}/\ell$ is a characteristic collision frequency.
Thus, the CE expansions \eqref{2} can be recast into the
dimensionless forms
\beq
P_{xy}/p=- \wa F_{xy}(\wa),\quad P_{xx}/p=1-\frac{4}{3}\mu
F_{xx}(\wa),\quad F_{xy}(\wa)\equiv\sum_{k=0}^\infty c_k
\wa^{2k},\quad F_{xx}(\wa)\equiv\sum_{k=0}^\infty c_k' \wa^k,
\label{13}
\eeq
where $c_k\equiv (\eta_k/p)\nu^{2k+1}$ and $c_k'\equiv
(\eta_k'/p)\nu^{k+1}$. Despite the simple definitions of the USF and
the ULF, no exact solution of the nonlinear Boltzmann equation is
known for those states. However, the problem becomes solvable in the
framework of the Bhatnagar--Gross--Krook (BGK) model kinetic
equation \cite{GS03,S00} and the solutions show that, except for
Maxwell molecules, the CE expansions \eqref{2} are \emph{divergent}
\cite{SBD86,S00}.

So far we have implicitly assumed  conventional gases made of
particles that collide \emph{elastically}. What about the case of
\emph{granular} gases? A granular gas is a large collection of
macroscopic particles which collide \emph{inelastically} and are
maintained in a fluidized state. It can be conveniently modeled as a
gas of smooth inelastic hard spheres characterized by a constant
coefficient of normal restitution $\alpha<1$ \cite{BP04}. For such a
model, the CE method has been applied to the inelastic Boltzmann
equation and the Navier--Stokes transport coefficients have been
derived \cite{BDKS98}. Taking into account that  the nonlinear
subclasses \eqref{2} of the full CE expansions, as said before,  do
not converge when the gas is made of elastic hard spheres
($\alpha=1$) \cite{SBD86,S00}, and considering the inherently
non-Newtonian nature of the steady USF of granular gases
\cite{SGD04}, it seems intuitive to expect that the  CE series
\eqref{2} are also divergent in the case of  inelastic hard spheres.
However, it turns out that, paradoxically, the series \eqref{2} do
converge in the inelastic case, the radii of convergence increasing
with increasing inelasticity \cite{S07,S08}.

\paragraph{Rheological properties}
In the USF, the mass and momentum balance equations yield constant
density $n$ and shear rate $\gd_{xy}$. However,
$n(t)=n(0)/[1+\gd_{xx}(0)t]$ and
$\gd_{xx}(t)=\gd_{xx}(0)/[1+\gd_{xx}(0)t]$ in the ULF, so that
$\gd_{xx}>0$ corresponds to an \emph{expansion} of the gas, while
$\gd_{xx}<0$ corresponds to a \emph{condensation} phenomenon
\cite{GK96,S00}. In either case the ratio $\gd_{xa}(t)/n(t)$ is
constant. As for the energy balance equation of inelastic hard
spheres in the USF and ULF, one has
\begin{equation}
\partial _{t}T(t)=-({2\gd_{xa}}/{3n}) P_{xa}(t) -\zeta(t) T(t),
\label{2.8}
\end{equation}
where $\zeta$ is the so-called cooling rate \cite{BP04}, which is
approximately given by $\zeta=\frac{5}{12}(1-\alpha^2)\nu_0$,
$\nu_0\propto n \sqrt{T}$ being an effective collision frequency for
elastic spheres. The cooling term on the right-hand side of Eq.\
\eqref{2.8} competes with the viscous heating term in the USF
($\gd_{xy}P_{xy}<0$) and in the ULF with $\gd_{xx}<0$ (note that
$P_{xx}>0$), so that, depending on the initial state, the
temperature either grows or decreases with time until a steady state
is eventually reached \cite{SGD04,S08,AS07}, except in the elastic
case ($\zeta=0$). On the other hand, in the ULF with $\gd_{xx}>0$
both terms on the right-hand side of Eq.\ \eqref{2.8} represent
cooling effects, so $T(t)\to 0$ and no steady state exists.

In order to analyze in detail the CE series \eqref{2}, it is
convenient to consider the following  BGK-like kinetic model
\cite{BDS99} of the inelastic Boltzmann equation:
\beq
(\partial_t
+\mathbf{v}\cdot\nabla)f=-\nu(f-f_0)+({\zeta}/{2})\partial_\mathbf{v}\cdot[(\mathbf{v}-\mathbf{u})f],
\label{n1}
\eeq
where $f$ is the velocity distribution function, $f_0$  is the local
version of the homogeneous cooling state distribution {\cite{BP04}},
and $\nu=\frac{1+\alpha }{2}\nu_0$ is an effective collision
frequency,
 so that $\wz\equiv
\zeta/\nu=\frac{5}{6}(1-\alpha)$. Taking moments in Eq.\ \eqref{n1}
one gets
\beq
\text{USF:} \left\{
\begin{array}{l}
\partial_t P_{xy}=-\gd_{xy}
P_{yy}-(\nu+\zeta)P_{xy},\\
\partial_t
P_{yy}=\nu p-(\nu+\zeta)P_{yy},
\end{array}
\right.\quad \text{ULF:}\,\,
\partial_t
P_{xx}=\nu p-(\nu+\zeta+3\gd_{xx})P_{xx},
\label{6}
\eeq
where $p=nT$. Equations \eqref{2.8} and \eqref{6} constitute a
closed set of equations for the evolution of $\{T,P_{xy},P_{yy}\}$
(USF) or $\{T,P_{xx}\}$ (ULF). To describe the non-Newtonian
hydrodynamic regime  we must focus on the nonlinear dependence of
the {scaled quantities $F_{xy}(\wa(t))\equiv -P_{xy}(t)/p(t)\wa(t)$
and $F_{xx}(\wa(t))\equiv -(3/4)[P_{xx}(t)/p(t)-1]/\wa(t)$ as
functions of $\wa(t)\equiv\gd_{xa}(t)/\nu(t)\propto 1/\sqrt{T(t)}$.
Elimination of time in favor of $\wa$
 in Eqs.\ \eqref{2.8} and \eqref{6} yields
a single second-order ordinary differential equation for
$F_{xy}(\wa)$ \cite{S07}} and a  single first-order ordinary
differential equation for $F_{xx}(\wa)$ \cite{S08}.

\paragraph{Results and discussion}

%%%%%%%%%%%%%%%%%%%%%%%%%%%%%%%%%%%%%%%%%%%%
%% Sample figure:
%%
%% The option [height=...] scales the picture to the given height,
%% without it it would be printed at its nominal size
%%%%%%%%%%%%%%%%%%%%%%%%%%%%%%%%%%%%%%%%%%%%

\begin{figure}
  \includegraphics[width=\columnwidth]{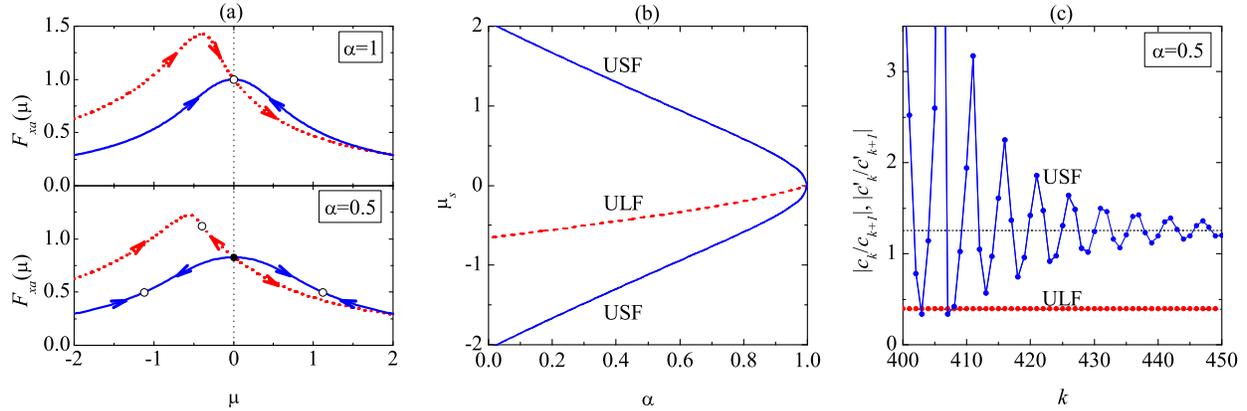}
\caption{(a) Plot of $F_{xy}(\wa)$ (solid lines) and $F_{xx}(\wa)$
(dotted lines) for $\alpha=1$ and $\alpha=0.5$; the open circles
denote the steady-state points. (b) Dependence of the steady-state
value $\wa_s$ on the coefficient of restitution $\alpha$. (c) Ratios
$|c_k/c_{k+1}|$  and $|c_k'/c_{k+1}'|$  for $\alpha=0.5$; the
horizontal dotted lines represent $\wa_s^2$ and $|\wa_s|$,
respectively.
\label{fig1}}
\end{figure}

The numerical solutions for a gas of elastic hard spheres
($\alpha=1$) and for a gas of highly inelastic hard spheres
($\alpha=0.5$) are plotted in Fig.\ \ref{fig1}(a). The arrows on the
curves indicate the direction followed by the time evolution of
$\wa(t)$. In the inelastic case ($\alpha<1$, $\wz\neq 0$) the
evolution leads to the steady-state point
$\wa_s=\pm\sqrt{3\wz/2}(1+\wz)$, $F_{xy}(\wa_s)=1/(1+\wz)^{2}$ for
the USF and $\wa_s=-3\wz(1+\wz)/2(1+3\wz)$,
$F_{xx}(\wa_s)=(1+3\wz)/(1+\wz)^{2}$ for the ULF (provided that
$\gd_{xx}<0$; otherwise, $\wa(t)\to \infty$).  On the other hand, in
the elastic case ($\alpha=1$, $\wz=0$) the temperature monotonically
increases with time in the USF and in the ULF with $\gd_{xx}<0$ and
thus the Knudsen number asymptotically vanishes, i.e., $\wa(t)\to
0$. The dependence of  $\wa_s$ on $\alpha$ is shown in Figure
\ref{fig1}(b).

According to the preceding discussion,  the zero Knudsen number
value ($\wa=0$) is a ``repeller'' of the time evolution of $\wa(t)$
both for USF and ULF in the inelastic case. For elastic collisions,
however, the state $\wa=0$ is an ``attractor'' of  $\wa(t)$ (except
in the ULF with $\gd_{xx}>0$). Expressed in other terms, the fixed
point $\wa=0$ is \emph{unstable} against any USF or ULF
perturbation, no matter how weak it is, if the particles are
\emph{inelastic}. On the other hand, $\wa=0$ is \emph{stable} for
\emph{elastic} collisions in the USF and in the ULF with
$\gd_{xx}<0$. Since the CE expansions \eqref{2} and \eqref{13} are
carried out about (and measure the departure from) the reference
homogeneous state ($\wa=0$), it follows that the CE series diverge
if $\lim_{t\to\infty}\wa(t) = 0$ (elastic case), whereas they
converge if $\wa(t)$ flees from $\wa=0$ (inelastic case). In the
latter situation, in addition, the radius of convergence must
coincide with the steady-state value $\wa_s$. Consequently, one must
have $\lim_{k\to\infty}|c_k/c_{k+1}|=\wa_s^{2}$ and
$\lim_{k\to\infty}|c_k'/c_{k+1}'|=|\wa_s|$. This expectation is
confirmed by an exact evaluation of the coefficients $c_k$ and
$c_k'$ for $k\leq 450$, as illustrated by Fig.\ \ref{fig1}(c) for
$\alpha=0.5$. In mathematical terms, the nonlinear viscosity
functions $F_{xa}(\wa)$ possess a singularity located at $\wa=0$ in
the elastic case and at $\wa=\wa_s\neq 0$ in the inelastic case.

To conclude, the paradoxical regularization by inelasticity of the
CE series \eqref{2} is directly related to the time evolution of the
Knudsen number $|\wa(t)|$, and hence of the temperature, Eq.\
\eqref{2.8}, and thus it is not an artifact of the kinetic model
\eqref{n1}. Also, the convergence or divergence of the series
\eqref{2} does not depend on whether the system actually is in the
USF, the ULF, or in any other state. The advantage of the USF and
ULF is that the full CE series of the shear and  normal stresses
reduce to the partial series \eqref{2}, thus allowing one to explore
their character in a detailed way.

%%%%%%%%%%%%%%%%%%%%%%%%%%%%%%%%%%%%%%%%%%%%%%%%
%% BACKMATTER
%%%%%%%%%%%%%%%%%%%%%%%%%%%%%%%%%%%%%%%%%%%%%%%%

\begin{theacknowledgments}
 This work has been supported by the
Ministerio de Educaci\'on y Ciencia (Spain) through Grant No.\
FIS2007--60977 (partially financed by FEDER funds) and by the Junta
de Extremadura (Spain) through Grant No.\ GRU08069.
\end{theacknowledgments}



\bibliographystyle{aipproc}   % if natbib is available

\end{document}